\documentclass{article}

\usepackage{microtype}
\usepackage{graphicx}
\usepackage{subfigure}
\usepackage{booktabs}

\usepackage{hyperref}
\usepackage{colortbl}

\usepackage[accepted]{icml2025}

\usepackage{amsmath}
\usepackage{amssymb}
\usepackage{mathtools}
\usepackage{amsthm}

\usepackage[capitalize,noabbrev]{cleveref}

\theoremstyle{plain}

\theoremstyle{definition}

\theoremstyle{remark}

\usepackage[textsize=tiny]{todonotes}
 
\icmltitlerunning{LoRA on Stellar Spectra Foundation Models}

\begin{document}

\twocolumn[
\icmltitle{Finetuning Stellar Spectra Foundation Models with LoRA}

\begin{icmlauthorlist}
\icmlauthor{Xiaosheng Zhao}{jhu}
\icmlauthor{Yuan-Sen Ting}{osu}
\icmlauthor{Alexander S. Szalay}{jhu}
\icmlauthor{Yang Huang}{ucas}
\end{icmlauthorlist}

\icmlaffiliation{jhu}{Department of Physics \& Astronomy, The Johns Hopkins University, Baltimore, MD 21218, USA}
\icmlaffiliation{osu}{Department of Astronomy, The Ohio State University, 140 West 18th Avenue, Columbus, OH 43210, USA}
\icmlaffiliation{ucas}{School of Astronomy and Space Science, University of Chinese Academy of Sciences, Beijing 100049, People's Republic of China}

\icmlcorrespondingauthor{Xiaosheng Zhao}{xzhao113@jh.edu}

\icmlkeywords{LoRA, Foundational Model, Stellar Spectra, Finetune}

\vskip 0.3in
]

\printAffiliationsAndNotice{}  
\begin{abstract}
Foundation models are beginning to impact stellar spectroscopy, where spectra encode rich physical information in a structured, language-like form. A key challenge is adapting these models across heterogeneous surveys with differing resolution and coverage. We apply Low-Rank Adaptation (LoRA) to fine-tune SpecCLIP—a contrastively pre-trained model on LAMOST and Gaia XP spectra—for downstream tasks on DESI Early Data Release (EDR) spectra. We show that LoRA enables few-shot learning on DESI, with performance varying by fine-tuned module and benefiting from Gaia XP knowledge embedded in the pre-trained model. Our results demonstrate that LoRA provides a lightweight and effective strategy for extending spectral foundation models to new instruments and survey domains.
\end{abstract}

\section{Introduction}
\label{sec: intro}

In recent decades, large-scale spectroscopic surveys have transformed our understanding of the Milky Way's formation and evolution. This progress has been driven by two parallel advances: the continual expansion of spectroscopic datasets and the development of increasingly sophisticated methods to infer stellar properties from spectra. These methods range from traditional template-matching algorithms like UlySS \citep{2009A&A...501.1269K} and LSP3 \citep{2015MNRAS.448..822X} to machine learning techniques such as The Cannon \citep{2015ApJ...808...16N}, The Payne \citep{2019ApJ...879...69T, 2019ApJS..245...34X, 2024ApJS..273...19Z} and TransformerPayne \citep{2025ApJ...980...66R}. Despite their success, most of these approaches rely on explicit supervision and are constrained either by limited empirical templates or imperfect synthetic models. The challenge becomes more severe when combining heterogeneous spectra from surveys with different wavelength coverages, resolutions, and signal-to-noise ratios, making it difficult to achieve consistent, survey-independent parameter estimates.

Motivated by the recent success of large language models (LLMs) in learning general-purpose representations from large-scale data, we previously introduced SpecCLIP \citep{2025arXiv250701939Z}, a spectral foundation model trained on LAMOST low-resolution spectra (LRS) and Gaia XP spectra using contrastive learning and some extensions to learn/arange both shared and non-shared information in the embeddings. Inspired by Contrastive Language-Image Pre-training (CLIP) \citep{2021arXiv210300020R} and AstroCLIP \citep{2024MNRAS.531.4990P}, the first introduction of CLIP into astronomy, SpecCLIP aligns representations across surveys and enables cross-modal applications such as similarity search, parameter estimation, and cross-modal prediction. The pre-trained models exhibit strong few-shot performance, and they provide a unified framework for homogenizing stellar parameters across instruments.

In this follow-up study, we address a critical challenge in foundation model deployment: how to adapt pre-trained models to new spectroscopic domains with minimal supervision. Specifically, we explore the use of Low-Rank Adaptation (LoRA, \citealp{2021arXiv210609685H}) to fine-tune different components of SpecCLIP on the DESI Early Data Release (EDR, \citealp{2024MNRAS.533.1012K}) stellar spectra. To our knowledge, this is the first application of LoRA in stellar spectroscopy. We find that LoRA enables effective few-shot learning on DESI data, even though the original model was trained exclusively on LAMOST and Gaia XP spectra. Our analysis shows that tuning different modules of the pre-trained model yields varying performance gains, and that Gaia XP information embedded in the foundation model can enhance adaptation to DESI, despite differences in resolution and wavelength coverage, and alignment with LAMOST spectra originally. Our approach facilitates the rapid deployment of powerful pre-trained models across heterogeneous spectroscopic datasets, enabling consistent stellar parameter estimation in the era of large, diverse sky surveys.

\begin{figure*}[ht]
\vskip 0.1in
\begin{center}
    \centerline{\includegraphics[width=0.85\textwidth]{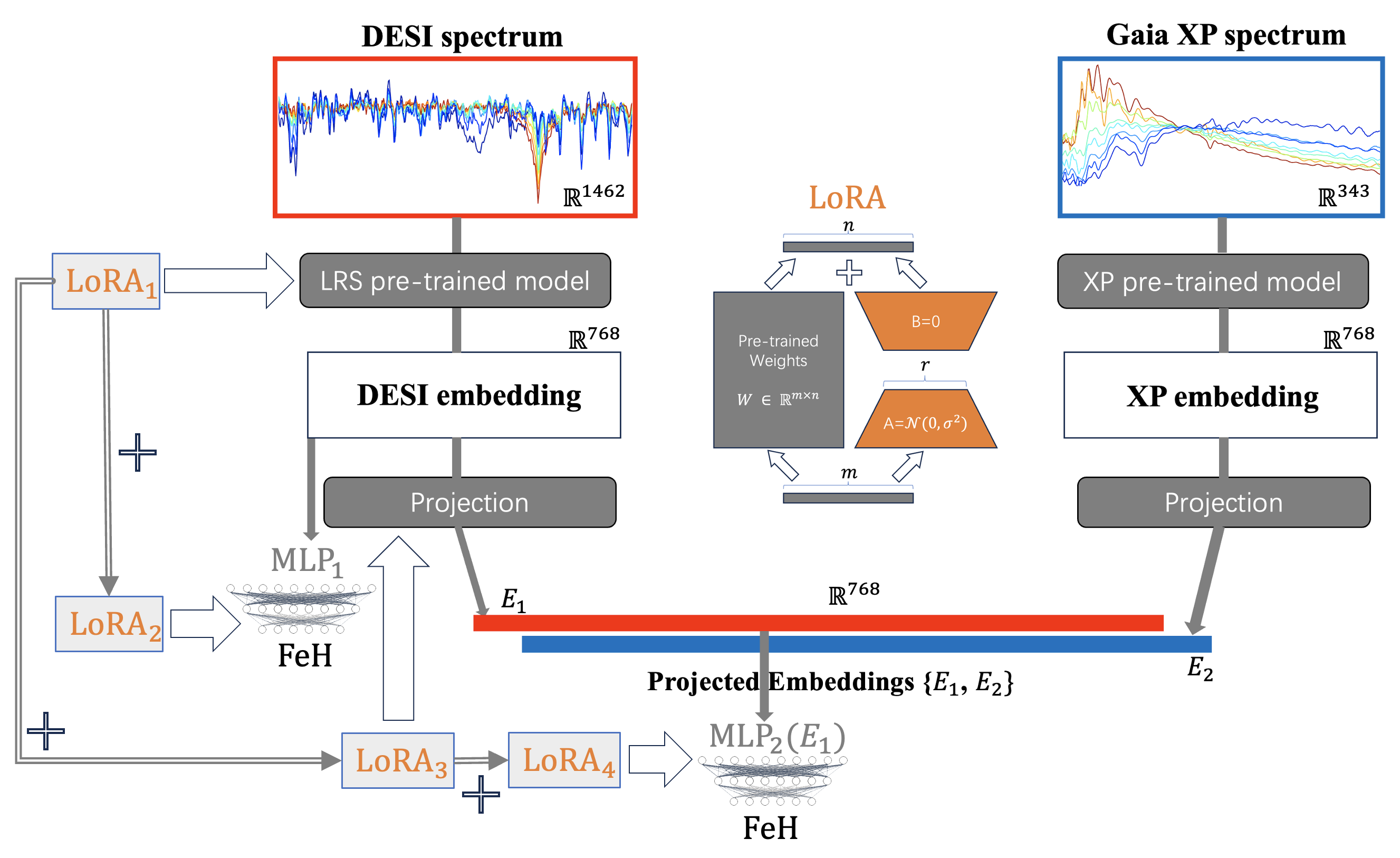}}
    \caption{
    Fine-tuning the SpecCLIP model with LoRA. Gray rounded blocks represent pre-trained modules: the LAMOST LRS and Gaia XP foundation models (trained separately on their respective modalities), and the projection networks trained with contrastive loss on paired projected embeddings. Two downstream MLPs, previously trained on LRS-based iron abundance labels, are also included. In the fine-tuning stage, a small number of DESI spectra are used to adapt: (1) the LRS-based MLP (\textbf{LoRA2}), (2) both the MLP and LRS foundation model (\textbf{LoRA1 + LoRA2}), (3) the projected-based MLP  (\textbf{LoRA4}), or (4) the projected-based MLP, projection network, and optionally the LRS foundation model (\textbf{LoRA1 + LoRA3 + LoRA4}). The center inset illustrates the LoRA mechanism, which updates the model weights by injecting the product of two low-rank matrices as a residual into the frozen pre-trained network.}

    \label{fig:LoRA_finetune}
\end{center}
\vskip -0.3in
\end{figure*}

\section{Method}
\label{sec:method}

\subsection{Overview of SpecCLIP Foundation Models}

SpecCLIP is a foundation model framework designed to align stellar spectra across modalities that differ in wavelength coverage, resolution, and signal-to-noise ratio. 

In the based SpecCLIP model, each modality—such as LAMOST LRS (DR11\footnote{https://www.lamost.org/dr11/}) and Gaia XP \citep{2023A&A...674A..33G} spectra—is assigned its own
transformer-based model with multi-head attention layers pre-trained using masked modeling \citep{2018arXiv181004805D} or multilayer perceptron (MLP)-based autoencoder. This unsupervised pre-training retrain the most important information which is further aligned by the contrastive training followed. It allows the resulting embeddings to support downstream tasks such as stellar parameter estimation, similarity search, and cross-modal prediction, with moderate supervision. 

\textbf{{LAMOST LRS Foundation Model}:} We select 966,082 ($9:1$ train-valid split, likewise, in the following cases) high-quality spectra from LAMOST DR11 for pre-training, applying quality cuts of G-band S/N$_g > 50$ and absolute g-band magnitude $< 15.8$. We restrict the wavelength range to 400--560~nm, yielding 1462 flux points sampled uniformly in logarithmic wavelength space per spectrum. These are normalized and segmented into overlapping tokens of size 20 with a stride of 10, resulting in 146 tokens per input. A transformer encoder with 6 layers processes these tokens to produce a spectral embedding. During pre-training, masked modeling is applied by masking 6 randomly selected chunks (each 20 tokens). 

\textbf{{Gaia XP Foundation Model}:}
We construct two types of models for Gaia XP spectra, which consist of 343 flux points spanning 336--1021~nm. For pre-training, we use 1,000,000 spectra. The input spectra are normalized by dividing by the V-band (550~nm) flux to provide a color-based inductive bias. Our model for Gaia XP spectra is an MLP-based autoencoder trained without masking. Both models (LRS and XP) share the same embedding dimension (768), number of parameters (42.7M), and number of training epochs.

\textbf{Contrastive Training:}
We use a dataset of 820,568 paired spectra from LAMOST LRS and Gaia XP, where each spectrum is processed by its corresponding pre-trained foundation model. The resulting embeddings are then mapped into a shared latent space using modality-specific projection networks. Each projection head (with 7.1M parameters) includes a cross-attention block, where a learnable query vector attends to the sequence of spectral features to produce a 768-length representation for contrastive alignment.

\textbf{Iron Abundance Estimation:}
We trained downstream MLP models to estimate stellar iron abundance ([Fe/H]) using 90,106 LAMOST LRS spectra, split into training and validation sets with a 9:1 ratio. The MLP takes as input either the spectral embeddings from the LAMOST LRS foundation model or the projected embeddings after contrastive alignment.

\subsection{LoRA Fine-Tuning for Cross-Survey Adaptation}
\label{sec:lora}

To adapt SpecCLIP to new spectroscopic domains with minimal supervision, we apply Low-Rank Adaptation (LoRA) to fine-tune the pre-trained models on stellar spectra from the DESI Early Data Release (EDR). LoRA introduces low-rank update matrices into pre-existing linear layers, allowing the original pre-trained weights to remain frozen while efficiently injecting task-specific information through a small number of additional parameters.

In essence, LoRA decomposes the weight update $\Delta W \in \mathbb{R}^{m \times n}$ into a product of two low-rank matrices: $\Delta W = A B$, where $A \in \mathbb{R}^{m \times r}$ and $B \in \mathbb{R}^{r \times n}$ for a given rank $r \ll \min(m, n)$. This enables substantial parameter savings while maintaining expressive adaptation capacity. In our setup, LoRA modules are inserted into selected attention and/or feed-forward layers of the pre-trained transformer backbones. A schematic of LoRA is shown in Fig.~\ref{fig:LoRA_finetune} (middle panel).

\textbf{Fine-Tuning with LoRA.} Our final model architecture is shown in Fig.~\ref{fig:LoRA_finetune}. We fine-tune four modules using Low-Rank Adaptation (LoRA): (1) the LAMOST-LRS foundation model (\textbf{LoRA1}), where LoRA is applied to all self-attention layers (\texttt{\{q,k,v,out\}\_proj}) with rank $=4$ and scaling factor $\alpha = 8$, adding 129{,}024 trainable parameters (0.30\% of the model); (2) the projection network used for contrastive alignment (\textbf{LoRA3}), where we apply LoRA to the \texttt{out\_proj} and subsequent linear layers with rank $=16$, $\alpha = 32$, and 147{,}456 trainable parameters (0.29\%); and (3) the downstream MLPs (\textbf{LoRA2} and \textbf{LoRA4}) trained on SpecCLIP embeddings, where LoRA is added to all linear layers with rank $=8$, $\alpha = 16$, and 31{,}752 parameters (2.30\% of the MLP size). We compare performance under different configurations and evaluate whether Gaia XP knowledge embedded in the pre-trained SpecCLIP model improves adaptation to DESI spectra without direct supervision. Each experiment completes in under 10 to 180 seconds on a single NVIDIA V100, depending on the model.

\section{Data}
\label{sec: data}

We base our experiments on the DESI Early Data Release (EDR) Milky Way Stars Value-Added Catalog, \texttt{mwsall-pix-fuji.fits}, which provides physical parameters and spectroscopic metadata for 625,588 sources. We extract stellar samples following a series of quality filters: we retain only rows where the spectral type is \texttt{STAR}, \texttt{RVS\_WARN} is zero, the observation is marked as \texttt{PRIMARY}, signal-to-noise ratio (SNR) exceeds 20, celestial coordinates are finite, and the Gaia \texttt{SOURCE\_ID} is valid (not \texttt{999999}). This results in a clean sample of DESI spectra suitable for foundation model adaptation. 

To provide ground-truth labels for downstream tasks, we cross-match this DESI sample with APOGEE DR17\citep{2022ApJS..259...35A}, yielding 495 stars with high-fidelity iron abundance ([Fe/H]) measurements. Among these, we randomly allocate 89 samples for training, 9 for validation, and the remaining 396 for testing. These are used to fine-tune \textbf{LoRA2} and \textbf{LoRA4}.

To fine-tune \textbf{LoRA1} and \textbf{LoRA3}, we exclude all 495 APOGEE-matched stars and randomly select an additional 164 unmatched DESI samples—86 for training and 78 for validation\footnote{Due to data preprocessing constraints, 86 training and 78 validation samples were ultimately used (intended: 100 each).}—with SNR further constrained to exceed 50. Although label-free, these spectra support contrastive learning and masked modeling objectives. While one could in principle use the same unlabeled DESI spectra for both foundation and downstream fine-tuning (due to the lack of label usage in the former), we deliberately use separate subsets in this study to make the adaptation task more challenging and test the generalization capacity of our method.

All DESI spectra, retrieved via the SPectra Analysis \& Retrievable Catalog Lab (SPARCL, \citealp{2024arXiv240105576J, 9347681}) at NOIRLab's Astro Data Lab \citep{10.1117/12.2057445, NIKUTTA2020100411}, are normalized using the same pipeline as for LAMOST LRS, and then interpolated onto the LAMOST wavelength grid spanning 400--560~nm. This standardization ensures compatibility between DESI and LAMOST representations.

For full SpecCLIP fine-tuning (combining LoRA1 and LoRA3), we further cross-match the selected DESI samples with the Gaia XP catalog to retrieve their fluxes. These are processed identically to the XP spectra in the original SpecCLIP paper: fluxes are divided by the value at 550~nm to yield a 343-dimensional color vector. 

\section{Results}
\label{sec: results}

\begin{table}[ht]
\centering
\caption{Performance of iron abundance ([Fe/H]) prediction on the DESI EDR  test set. We report robust standard deviation ($\sigma$, using Tukey Biweight Scale Estimator\citep{hoaglin1983understanding}) of the residual and $R^2$ scores for different fine-tuning configurations. Results are shown for the full test set and for two iron abundance subsets. Top three entries are highlighted with decreasing shades of blue. Numbers in brackets indicate the source counts.}
\label{tab:results}
\begin{tabular}{lcc}
\toprule
\textbf{Method} & $\sigma$ & $R^2$ \\
\midrule
\multicolumn{3}{l}{\textit{Full Test Set (396)}} \\
Zero-shot (MLP1) & 0.2730 & 0.7358 \\
LoRA2            & 0.2663 & 0.7156 \\
LoRA1 + LoRA2    & \cellcolor{blue!12}0.2227 & \cellcolor{blue!6}0.7719 \\
Zero-shot (MLP2) & 0.2560 & 0.7203 \\
LoRA4            & \cellcolor{blue!20}0.2023 & \cellcolor{blue!20}0.7937 \\
LoRA1 + LoRA3 + LoRA4 & \cellcolor{blue!6}0.2297 & \cellcolor{blue!12}0.7801 \\
\midrule
\multicolumn{3}{l}{\textit{[Fe/H] $<$ –1 (60, metal-poor)}} \\
Zero-shot (MLP1) & \cellcolor{blue!20}0.4444 & \cellcolor{blue!20}-0.5130 \\
LoRA2            & 0.5872 & -1.2881 \\
LoRA1 + LoRA2    & \cellcolor{blue!6}0.5151 & -0.9143 \\
Zero-shot (MLP2) & \cellcolor{blue!12}0.4658 & \cellcolor{blue!12}-0.7579 \\
LoRA4            & 0.5803 & -0.8357 \\
LoRA1 + LoRA3 + LoRA4 & 0.5970 & \cellcolor{blue!6}-0.8159 \\
\midrule
\multicolumn{3}{l}{\textit{[Fe/H] $\geq$ –1 (336, metal-rich)}} \\
Zero-shot (MLP1) & 0.2479 & 0.0702 \\
LoRA2            & 0.2272 & 0.2378 \\
LoRA1 + LoRA2    & \cellcolor{blue!6}0.1924 & \cellcolor{blue!6}0.4173 \\
Zero-shot (MLP2) & 0.2371 & 0.0725 \\
LoRA4            & \cellcolor{blue!20}0.1621 & \cellcolor{blue!20}0.5106 \\
LoRA1 + LoRA3 + LoRA4 & \cellcolor{blue!12}0.1851 & \cellcolor{blue!12}0.4274 \\
\bottomrule
\end{tabular}
\vskip -0.3in
\end{table}

\begin{figure}[ht]
\vskip 0.05in
\begin{center}
    \centerline{\includegraphics[width=\columnwidth]{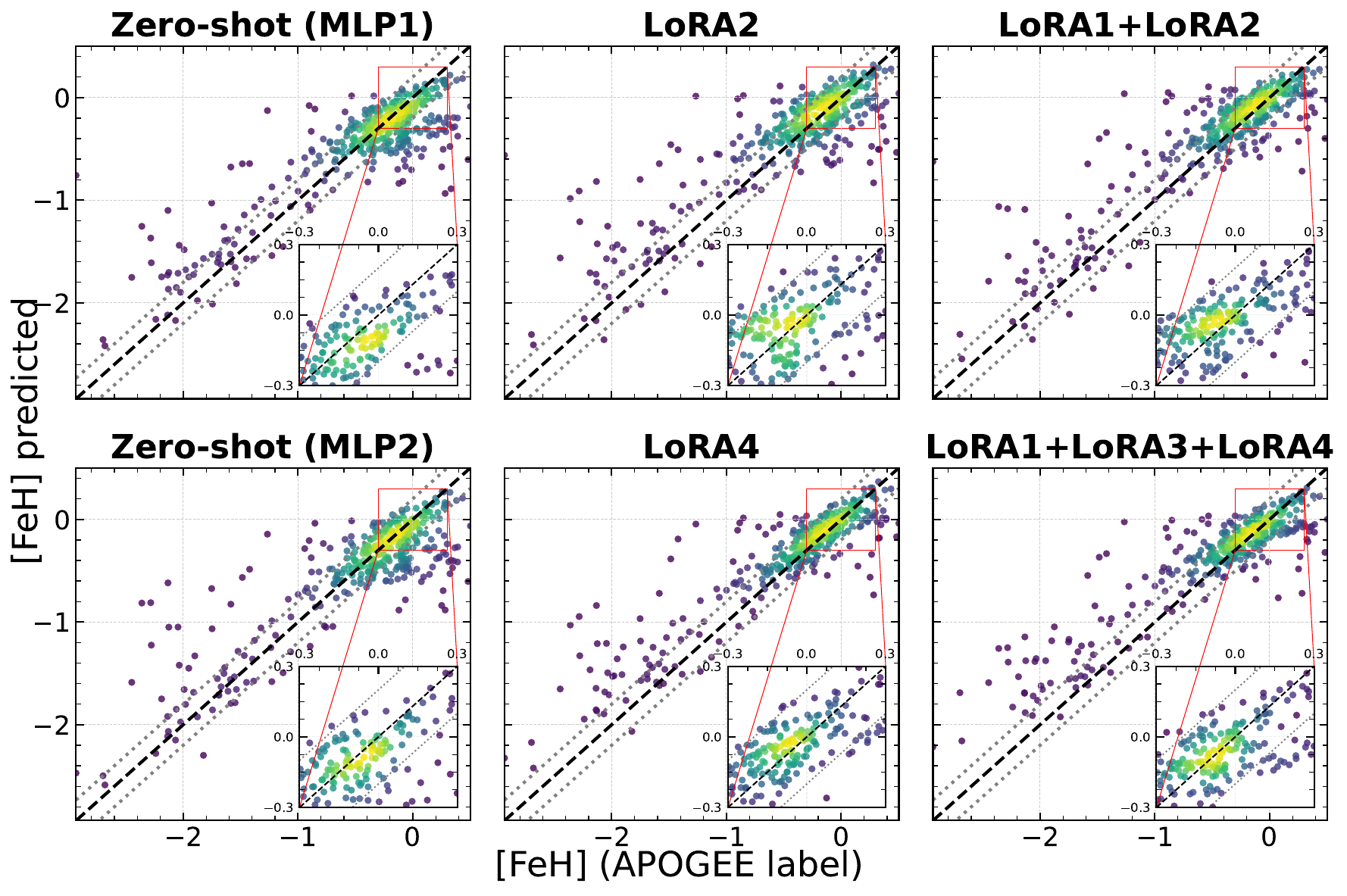}}
    \caption{Scatter plots of predicted versus APOGEE iron abundance ([Fe/H]) for all fine-tuning configurations, with point density indicated by the \texttt{viridis} colormap. The dashed black line indicates the ideal 1:1 relation, and the gray dotted lines mark ±0.2 dex deviations. An inset zooms into the $\pm$0.3 dex region for closer inspection. LoRA fine-tuning improves overall predictive accuracy compared to zero-shot baselines, particularly in the metal-rich regime ([Fe/H] $\geq -1$). However, performance in the metal-poor regime ([Fe/H] $< -1$) remains limited, with larger scatter and systematic deviations across all configurations.}
    \label{fig:LoRA_finetune_scatter}
\end{center}
\vskip -0.4in
\end{figure}

We evaluate the effectiveness of LoRA-based fine-tuning under several configurations, comparing zero-shot performance with targeted adaptations to different parts of the SpecCLIP architecture. Table~\ref{tab:results} summarizes the robust standard deviation ($\sigma$) and coefficient of determination ($R^2$) on the DESI EDR test set, both overall and for subsets divided by iron abundance. Fig.\ref{fig:LoRA_finetune_scatter} shows the corresponding scatter plots for all experiments.

\textbf{Full test set.} The best overall performance is achieved by LoRA4, which fine-tunes only the downstream MLP for the Gaia XP-aligned embeddings, reaching the lowest robust scatter ($\sigma = 0.2023$) and the highest $R^2 = 0.7937$. Close behind is the full multi-module adaptation (LoRA1 + LoRA3 + LoRA4), showing consistent improvement over zero-shot baselines. Notably, LoRA1 + LoRA2 also performs well, indicating that the foundation model adaptations contribute meaningfully.

\textbf{Metal-poor subset ([Fe/H] $< -1$).} All methods struggle in this regime. Interestingly, the best performance comes from the zero-shot MLP1 model ($\sigma = 0.4444$, $R^2 = -0.5130$), suggesting that fine-tuning may overfit or underperform in this sparse label regime. All LoRA-enhanced variants yield higher $\sigma$ and more negative $R^2$, particularly LoRA2 and LoRA1 + LoRA3 + LoRA4, which degrade the predictive performance ($\sigma \sim 0.58$--$0.60$, $R^2 < -0.8$). This highlights the difficulty of generalizing to metal-poor stars with limited data.

\textbf{Metal-rich subset ([Fe/H] $\geq -1$).} In contrast, all fine-tuned models significantly outperform the zero-shot baselines. LoRA4 achieves the best results ($\sigma = 0.1621$, $R^2 = 0.5106$), with LoRA1 + LoRA3 + LoRA4 and LoRA1 + LoRA2 close behind. This suggests that LoRA fine-tuning is highly effective for the dominant population of metal-rich stars in DESI, where the training distribution is better matched.

Overall, these results demonstrate that LoRA offers a flexible and lightweight way to adapt foundation models to new survey domains. Fine-tuning the downstream MLP alone (LoRA2 or LoRA4) is effective, but further gains can be achieved by jointly adapting the foundation model. However, the poor performance in the metal-poor regime suggests the need for better regularization or targeted augmentation to ensure robust generalization in low-density regions of label space.

\section{Summary}
\label{sec: summary}

We presented the first application of Low-Rank Adaptation (LoRA) to fine-tune a spectral foundation model for stellar spectroscopy. Building on the SpecCLIP framework, we explored how LoRA can efficiently adapt different modules--pre-trained LRS foundation model, projection heads, and downstream MLPs--to the DESI EDR stellar spectra with only a few (around 100 to 200) labeled examples.

Our experiments demonstrate that LoRA fine-tuning leads to noticeable improvements over zero-shot baselines, particularly in the metal-rich regime. We find that even lightweight adaptation of the downstream predictor achieves strong performance, while further gains can be achieved by jointly tuning the foundation model layers. However, performance on metal-poor stars remains limited, suggesting that additional strategies—such as better regularization, improved sample balancing for fine-tuning, careful hyperparameter optimization, or uncertainty-aware modeling—may be necessary to enhance generalization in this regime.

These findings suggest that LoRA provides a scalable and effective pathway for extending pre-trained spectral foundation models to new surveys, enabling flexible and data-efficient inference in heterogeneous spectroscopic environments. 

\section*{Broader Impact}

This paper presents work whose goal is to advance the field of Machine Learning in the context of astrophysical data analysis. There are many potential societal consequences of our work, none which we feel must be specifically highlighted here.

\section*{Acknowledgements}
We acknowledge support through the generosity of Eric and Wendy Schmidt, by recommendation of the Schmidt Futures program. Y.S.T is supported by the National Science Foundation under Grant No. AST-2406729. Y.H. acknowledges the supported from the National Science Foundation of China (NSFC grant No. 12422303). This research uses services or data provided by the SPectra Analysis and Retrievable Catalog Lab (SPARCL) and the Astro Data Lab, which are both part of the Community Science and Data Center (CSDC) Program of NSF NOIRLab. NOIRLab is operated by the Association of Universities for Research in Astronomy (AURA), Inc. under a cooperative agreement with the U.S. National Science Foundation.

\bibliography{main}
\bibliographystyle{icml2025}

\newpage
\appendix
\onecolumn

\end{document}